\documentclass[11pt,a4paper]{article}

\usepackage[a4paper,margin=2.35cm]{geometry}
\usepackage[T1]{fontenc}
\usepackage[utf8]{inputenc}
\usepackage{lmodern}
\usepackage{microtype}
\usepackage{graphicx}
\usepackage{booktabs}
\usepackage{tabularx}
\usepackage{array}
\usepackage{enumitem}
\usepackage[table]{xcolor}
\usepackage{hyperref}
\usepackage[numbers,sort&compress]{natbib}

\hypersetup{
  colorlinks=true,
  linkcolor=blue!55!black,
  citecolor=blue!55!black,
  urlcolor=blue!55!black,
  pdftitle={SotA Lens: A Network-Augmented Methodology and Tool for Exploratory State-of-the-Art Reviews},
  pdfauthor={Diogo Peralta Cordeiro}
}

\newcommand{\tool}{\emph{SotA Lens}}
\newcommand{\sota}{State-of-the-Art}
\newcommand{\dpm}{Dynamic Projection-Mapping}
\newcommand{\sar}{Spatial Augmented Reality}
\newcommand{\pmapping}{Projection Mapping}
\newcommand{\rttpm}{Real-time Tracking Projection Mapping}
\newcommand{\procam}{projector--camera}

\title{\tool: A Network-Augmented Methodology and Tool for Exploratory \sota{} Reviews}

\author{
Diogo Peralta Cordeiro\\
Faculty of Engineering, University of Porto, Portugal\\
\href{mailto:mail@diogo.site}{mail@diogo.site}\\
\href{https://orcid.org/0000-0002-0260-5121}{ORCID: 0000-0002-0260-5121}
}

\date{Preprint prepared 5 May 2026}

\begin{document}
\maketitle

\begin{abstract}
Researchers often begin new projects by conducting a broad \sota{} review before they are ready to define the narrow protocol required by a systematic review. This is especially common in multidisciplinary areas where terminology is unstable, communities are weakly connected, and relevant work is dispersed across technical and application domains. This paper presents \tool, a network-augmented methodology and lightweight software toolkit for exploratory \sota{} reviews. The approach combines documented seed search, DOI-level metadata resolution, bounded citation expansion, directed graph construction, community detection, ranking of authors and subject terms, and human labelling of research communities. It is designed to complement, not replace, established review protocols such as PRISMA, PRISMA-ScR, systematic mapping studies, and bibliometric science mapping. The method is demonstrated through a proof-of-concept review of \dpm{} and \sar{}. Starting from approximately 200 seed search results, the workflow produced a citation graph with 2,198 DOI-level vertices and 8,249 reference edges; a filtered largest component for 2010--2023 contained 986 vertices, 2,693 edges, and sixteen labelled communities. The contribution is both methodological and practical: \tool{} helps researchers map broad fields, identify clusters and gaps, and produce auditable review artifacts before committing to a narrower systematic review protocol. This paper is not intended as a domain survey of \dpm{} or \sar{}; rather, it introduces and demonstrates an original review-support methodology and software artifact using that domain as a proof-of-concept case study.
\end{abstract}

\noindent\textbf{Keywords:} state-of-the-art review; literature review methodology; citation graph; network science; bibliometrics; scoping review; systematic mapping; research software

\section{Introduction}

Literature reviews serve different purposes. Some reviews answer narrowly specified empirical questions; others map concepts, methods, communities, and gaps in a research area. This distinction matters because the methodological requirements of an intervention-focused systematic review are not identical to those of an exploratory \sota{} review. In early doctoral research, design science, and emerging technical fields, the immediate problem is often not to aggregate effect sizes or decide whether an intervention works. Instead, the researcher must first understand what a field is, what neighbouring terms it uses, which communities produce relevant work, and which technical or conceptual gaps should structure the next stage of inquiry.

This paper addresses that earlier exploratory stage. It proposes \tool, a network-augmented methodology and software toolkit for building and inspecting literature corpora as citation graphs. The approach was originally developed to support a broad review of \dpm{}, a field that overlaps with \sar{}, \pmapping{}, \procam{} systems, real-time tracking, interactive systems, museums, manufacturing, education, accessibility, and performance. Such a field is difficult to review by keyword strings alone because relevant work appears under several labels and in several venues. It is also difficult to review using a purely informal narrative approach because the breadth of the domain makes it easy to miss neighbouring communities.

Projection-based augmented reality has a long technical history in \sar{} and \procam{} systems~\citep{bimber2005sar,bimber2008procam}. More recent work in projection mapping and dynamic projection has consolidated many of the relevant calibration, geometric compensation, and deployment challenges~\citep{grundhofer2018projection,iwai2024projection}. However, for a researcher entering such a broad area, the first methodological difficulty is often not yet how to evaluate a particular system, but how to identify the structure of the surrounding literature. \tool{} addresses this practical problem.

The central claim is one of complementarity. \tool{} does not replace PRISMA-style systematic review, scoping-review protocols, systematic mapping, or mature bibliometric science mapping. Instead, it sits upstream of them. It provides structure and auditability at the moment when a researcher is still discovering the perimeter of a field. The contribution of this paper is threefold:
\begin{enumerate}[leftmargin=*]
    \item a reproducible methodology for exploratory \sota{} reviews based on seed retrieval, metadata resolution, bounded citation expansion, graph construction, community detection, and human synthesis;
    \item a lightweight software package consisting of a command-line data pipeline and a static web explorer; and
    \item a proof-of-concept case study showing how the method maps the \dpm{}/\sar{} literature into interpretable technical and application communities.
\end{enumerate}

The reusable contribution is the review workflow and software artifact, not a definitive survey of \dpm{} or \sar{}. The case study is used because it provides a real, difficult, multidisciplinary corpus on which to demonstrate the method.

\section{Background: review methods and their roles}

The term ``literature review'' covers a family of methods rather than a single procedure. Grant and Booth's typology emphasizes that review types differ in their search, appraisal, synthesis, and analysis commitments~\citep{grant2009typology}. This distinction is useful because no single review format is optimal for all research stages.

\paragraph{Narrative and critical reviews.}
Narrative and critical reviews are flexible and can synthesize ideas across fields, but their search and inclusion decisions are often hard to audit. They are appropriate when the goal is conceptual argument, but weaker when the reader must assess whether a search was comprehensive or whether relevant communities were missed.

\paragraph{Systematic reviews and PRISMA.}
Systematic reviews are strongest when the research question is narrow enough to define eligibility criteria, screening procedures, quality appraisal, and synthesis plans before the review is conducted. PRISMA 2020 improves the reporting of such reviews through structured items and flow diagrams~\citep{page2021prisma,prisma2020web}. However, PRISMA is primarily a reporting guideline rather than a discovery algorithm. It does not by itself solve the early-stage problem of finding the boundaries of a scattered field.

\paragraph{Scoping reviews and PRISMA-ScR.}
Scoping reviews are closer to the problem addressed here because they map broad topics and concepts. PRISMA-ScR provides reporting items for scoping reviews and evidence maps~\citep{tricco2018prisma,prismascrweb}. Yet scoping reviews still typically require explicit review questions, eligibility decisions, and charting procedures. \tool{} can be used upstream of a scoping review to help define the conceptual perimeter, vocabulary, and candidate subfields before a formal scoping protocol is fixed.

\paragraph{Systematic mapping studies.}
Systematic mapping studies classify a body of literature to reveal frequencies, venues, methods, and research gaps. They are particularly influential in software engineering~\citep{petersen2015mapping}. \tool{} shares this mapping orientation, but it emphasizes citation-network structure and exploratory community discovery before a final classification scheme is fixed.

\paragraph{Bibliometric science mapping.}
Bibliometric tools construct and visualize networks of authors, journals, citations, co-citations, bibliographic coupling, or term co-occurrence. VOSviewer is explicitly designed to construct and visualize bibliometric networks~\citep{van2010vosviewer,vosviewerweb}; CiteSpace emphasizes the detection of trends and transient patterns~\citep{chen2006citespace}; bibliometrix provides a comprehensive R ecosystem for bibliometric analysis~\citep{aria2017bibliometrix}. \tool{} is not intended to replace such mature tools. Its niche is smaller and more practical: a review-workflow bridge that transforms a corpus into auditable tables, graph artifacts, and a static browser-based explorer that can be hosted or archived with a paper.

\begin{table}[htbp]
\centering
\small
\begin{tabularx}{\textwidth}{p{2.85cm}XXX}
\toprule
\textbf{Approach} & \textbf{Primary role} & \textbf{Best stage} & \textbf{Main limitation addressed by \tool{}} \\
\midrule
Narrative / critical review & Conceptual synthesis across bodies of work & Early to mature, depending on purpose & Low auditability of search and boundary choices \\
PRISMA systematic review & Transparent answer to a narrow question & Mature review question & Too narrow for initial field discovery \\
PRISMA-ScR scoping review & Map evidence and concepts under a protocol & Broad but defined scope & Still requires a stable perimeter and charting plan \\
Systematic mapping & Classify a corpus by a scheme & Classification-ready corpus & Classification can be premature before communities are known \\
Bibliometric science mapping & Quantify and visualize scientific networks & Database-ready quantitative analysis & Can become metric-driven without review narrative \\
\tool{} & Explore a broad field using citation-network artifacts & Upstream exploratory \sota{} and research framing & Does not replace final screening, appraisal, or synthesis \\
\bottomrule
\end{tabularx}
\caption{Positioning of \tool{} relative to common review methodologies.}
\label{tab:comparison}
\end{table}

\section{Methodology: a network-augmented \sota{} review}

The proposed methodology combines structured search documentation, bounded citation expansion, network analysis, and human interpretation. It is not a fully automatic review method. The graph suggests structure; the researcher interprets that structure. The workflow has seven stages.

\subsection{Stage 0: define the exploratory perimeter}

\begin{figure}[htbp]
\centering
\includegraphics[width=.82\linewidth]{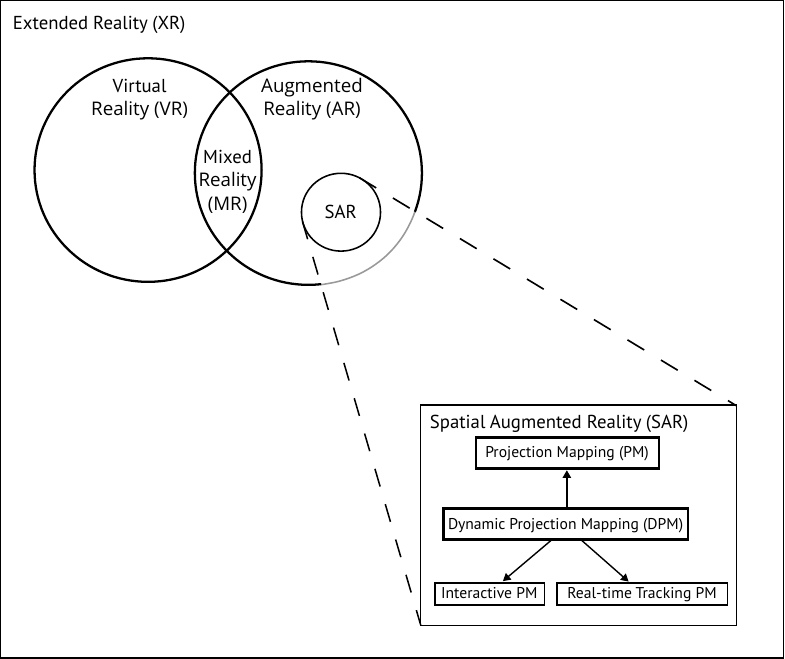}
\caption{Initial conceptual map of \dpm{} within neighbouring extended-reality and projection-mapping fields. This map was used to avoid prematurely reducing the review to a single keyword.}
\label{fig:fieldmap}
\end{figure}

The researcher begins by defining a broad field of inquiry and neighbouring terms. Unlike a systematic review protocol, this stage does not require a final eligibility scheme. Instead, it documents the initial assumptions: seed terms, neighbouring domains, expected application areas, and known terminology conflicts. In the proof-of-concept case, the primary field of inquiry was \dpm{}, but several neighbouring and overlapping areas were relevant: \sar{}, \pmapping{}, Dynamic Vision Systems, Dynamic Image Control, and \rttpm{} (Fig.~\ref{fig:fieldmap}).

\subsection{Stage 1: collect seed results}

Seed results are retrieved from one or more search sources. In the original \dpm{}/\sar{} case, the seed query was:
\begin{quote}
\emph{``Dynamic Projection Mapping'' OR ``Spatial Augmented Reality''.}
\end{quote}
The seed set should be recorded as a table containing at least title, source, and URL. DOI, year, authors, venue, and references can be resolved later. This separation is important because search interfaces and metadata services have different strengths.

\subsection{Stage 2: resolve metadata}

The seed table is resolved to DOI-level records through Crossref or a similar metadata service. For each article, the pipeline records DOI, title, authors, year, URL, subject terms, affiliations when available, and references. Crossref is a widely used source of community-owned scholarly metadata~\citep{hendricks2020crossref}. Failed resolutions are preserved rather than silently discarded. This creates an audit trail and makes coverage limitations visible.

\subsection{Stage 3: bounded citation expansion}

The method expands the corpus by retrieving references from seed papers. Expansion is bounded by a maximum depth. In the case study, depth 1 was sufficient to expose major communities while avoiding an explosion into tens of thousands of records. Deeper expansion is possible, but the cost is interpretability: each depth level increases the probability of drifting into generic background literature.

\subsection{Stage 4: graph construction}

The corpus becomes a directed graph $G=(V,E)$ where each vertex is a DOI-level article and each directed edge represents a reference relation from a citing article to a cited article. This representation makes the review landscape inspectable through degree, connected components, local neighbourhoods, and community structure.

\subsection{Stage 5: community detection and ranking}

The graph is analysed using community detection. The prototype used modularity-oriented community detection, a common approach for identifying dense groups in networks~\citep{newman2006modularity}. \tool{} also computes author and subject rankings. These rankings should not be interpreted as definitive bibliometric authority because author-name and affiliation disambiguation remain imperfect. Their purpose is practical: to help the reviewer notice recurring names, institutions, and domains.

\subsection{Stage 6: human labelling and narrative synthesis}

Detected communities are not self-explanatory. The researcher inspects representative papers in each cluster and assigns semantic labels. This stage is where the method deliberately combines computation and expert judgment. The graph suggests structure; the reviewer decides whether that structure corresponds to a method, application area, historical period, or artifact of the search.

\subsection{Stage 7: report artifacts}

A network-augmented \sota{} review should report the seed queries and search sources; corpus size and expansion depth; failed resolutions and known exclusions; graph size, filtering decisions, and community-detection method; ranked authors, subject terms, and/or institutions, with caveats; human-assigned community labels; and the final narrative synthesis. The aim is not simply to produce a graph image, but to leave behind auditable review artifacts.

\begin{figure}[htbp]
\small
\centering
\fbox{\begin{minipage}{.93\linewidth}
\textbf{Simplified article-search workflow.}
\begin{enumerate}[leftmargin=*,itemsep=2pt]
  \item Retrieve seed results for a broad subject query.
  \item For each seed article, resolve DOI, title, authors, year, URL, and references through metadata services.
  \item If the article depth is below the chosen maximum, resolve its references and append them to the exploration queue.
  \item Build a directed graph in which vertices are articles and edges are reference relations.
  \item Detect communities, compute rankings, and export corpus, graph, and web artifacts.
\end{enumerate}
\end{minipage}}
\caption{Simplified article-search and citation-expansion workflow implemented by \tool{}.}
\label{fig:workflow}
\end{figure}

\section{Software: \tool}

\tool{} operationalizes the above method in two components: a command-line data pipeline and a static web explorer.

\subsection{Command-line pipeline}

The command-line tool converts article tables into graph and web artifacts. It reads article metadata from CSV, normalizes DOI-based identifiers, extracts reference DOIs from common list-like encodings, constructs directed citation graphs with NetworkX~\citep{hagberg2008networkx}, computes summary statistics, and exports both Gephi-compatible GEXF files~\citep{bastian2009gephi} and browser-ready JSON. The pipeline deliberately treats metadata resolution, graph construction, and web export as separate stages so that intermediate artifacts can be inspected, corrected, and versioned.

The distributed version avoids live Google Scholar scraping. This is a reproducibility and maintenance decision: search-interface scraping is brittle, difficult to reproduce, and may conflict with service restrictions. Instead, the tool supports imported seed tables, local/offline corpora, Crossref-style metadata workflows, and graph artifacts generated outside the browser.

\subsection{Static web explorer}

The static web explorer is designed for simple deployment. It requires no database, server-side runtime, or build system. It can be hosted on GitHub Pages or any static website. Users can load local CSV and GEXF files directly in the browser, or explicitly open the bundled case study. The interface provides full-text search over titles, DOIs, authors, and subject terms; filters by year, depth, and degree; a canvas-based graph view; article inspection; rankings; community-size summaries; and CSV export of filtered results.

This design supports two forms of reproducibility. First, the repository can archive the corpus, graph, and preprint source. Second, readers can inspect the case-study graph in the browser without installing a specialized bibliometric environment.

\section{Proof of concept: \dpm{} and \sar{}}

The proof-of-concept case study used \dpm{}/\sar{} because it is a suitable stress test for exploratory review methods. The field has overlapping names, multiple technical subproblems, industrial and academic contributions, and several application domains. It sits at the intersection of projection mapping, \sar{}, \procam{} systems, real-time tracking, dynamic image control, interactive systems, and venue-specific applications.

\subsection{Corpus and expansion}

The initial search used the query ``Dynamic Projection Mapping'' OR ``Spatial Augmented Reality''. The first ten pages of search results for the relevant terms produced approximately 200 unique seed results. After metadata resolution and bounded reference expansion, the final DOI-level graph contained 2,198 vertices and 8,249 directed reference edges. The corpus covered publications from 1936 to 2023, with the primary interpretive focus on the period from 2010 to 2023. A further tertiary expansion was deliberately avoided because it would have increased the graph to orders of magnitude beyond what was useful for the review.

For auditability, the seed-search log records the search source, query string, retrieval date, number of raw results, number of deduplicated seed records, number of DOI-resolved records, and number of unresolved records where available. These intermediate tables are preserved as review artifacts rather than silently discarded.

The academic search was complemented by an industrial scan. In the original review, companies such as The Walt Disney Company, Dataton AB, Christie Digital Systems, Panasonic, and Sony were examined because projection mapping and \procam{} systems are shaped by commercial calibration, show-control, and deployment tools as well as academic prototypes. This industrial material was treated as contextual evidence rather than as part of the DOI citation graph.

\subsection{Graph and communities}

\begin{figure}[htbp]
\centering
\includegraphics[width=\linewidth]{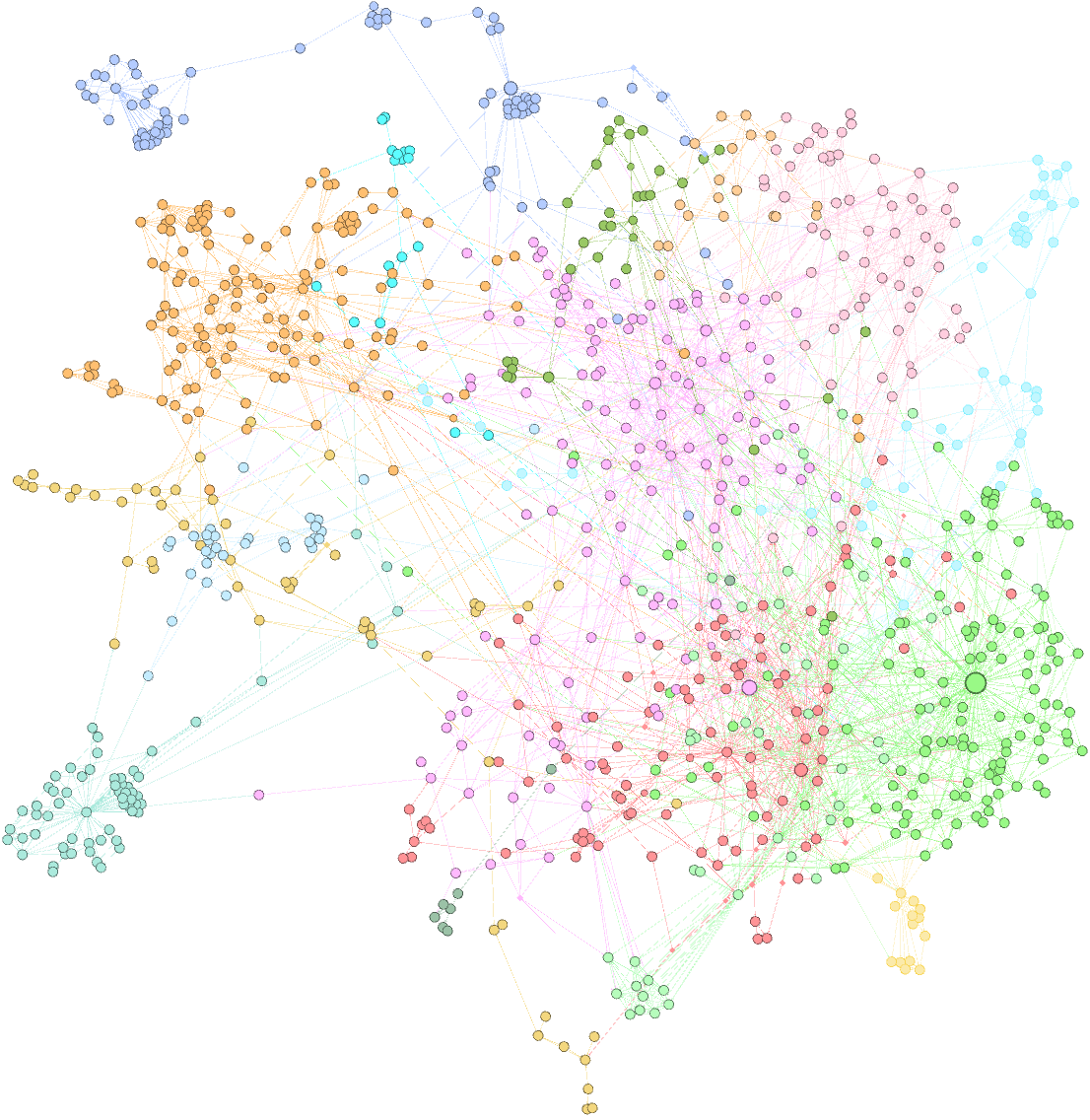}
\caption{Largest connected component of the 2010--2023 filtered citation graph. Each vertex is a DOI-level article; each directed edge is a reference relation; vertex size represents total degree; colour represents modularity class/community. The layout is used as an exploratory orientation artifact rather than as a statistical result.}
\label{fig:graph}
\end{figure}

Figure~\ref{fig:graph} shows the largest component of the subgraph corresponding to the period 2010--2023. This filtered component contains 986 vertices and 2,693 edges, representing approximately 42\% of the complete graph. Each vertex is a DOI-level article; each directed edge is a reference relation; vertex size represents total degree; and colour represents modularity class/community. The layout is used as an exploratory orientation artifact rather than as a statistical result. Table~\ref{tab:communities} provides the human-assigned labels used to interpret the coloured communities in the graph.

The analysis exposed recurring technical communities: projector calibration; Dynamic Vision Systems; Dynamic Image Control; marker-based and markerless \rttpm{}; interaction in projection mapping; projection onto non-rigid or flexible materials; human perception; and human cues in mixed and extended reality. It also exposed application communities: teleconferencing, multi-surface displays, food, nature and living things, projection onto humans, education, creative offices and architecture, exhibitions and museums, entertainment, medicine, manufacturing, robots, accessibility, and virtual try-on.

Table~\ref{tab:communities} summarizes the human-labelled categories that were most useful for narrative synthesis. The important point is not that these labels are definitive. They are a scaffold for reading: they show which parts of the literature require focused inspection.

\begin{table}[htbp]
\centering
\small
\caption{Human labels assigned to the modularity communities shown in Fig.~\ref{fig:graph}. Colours are used only as visual cues for interpreting the graph; the substantive categories are the community labels.}
\label{tab:communities}
\begin{tabularx}{\textwidth}{p{1.3cm}p{2.8cm}X}
\toprule
\textbf{ID} & \textbf{Visual cue} & \textbf{Human-assigned community label} \\
\midrule
\rowcolor[HTML]{A2DED2}
C1 & Eggshell blue & Projection mapping in health \\
\rowcolor[HTML]{F2B0F0}
C2 & Light lavender & Dynamic Projection-Mapping \\
\rowcolor[HTML]{A9CA83}
C3 & Light grey green & Interactive Projection Mapping \\
\rowcolor[HTML]{C3D5C8}
C4 & Light grey & Projection mapping in dance and performance \\
\rowcolor[HTML]{CED9F2}
C5 & Light blue grey & Dynamic Projection-Mapping in creative contexts \\
\rowcolor[HTML]{D6E9F2}
C6 & Cloudy blue & Augmented and Spatial Augmented Reality in health \\
\rowcolor[HTML]{F2DAE2}
C7 & Very light pink & Dynamic Image Control \\
\rowcolor[HTML]{ECDFB4}
C8 & Pale yellow & Extended reality in education \\
\rowcolor[HTML]{F2D4AE}
C9 & Light peach & Extended reality in manufacturing and industrial assembly \\
\rowcolor[HTML]{A5F1F2}
C10 & Pale cyan & Spatial Augmented Reality with robots \\
\rowcolor[HTML]{F2BFC1}
C11 & Light rose & Dynamic Vision Systems \\
\rowcolor[HTML]{91EC7F}
C12 & Pastel green & Projection mapping onto human faces and bodies \\
\rowcolor[HTML]{C2EFB8}
C13 & Light sage & Projector calibration \\
\rowcolor[HTML]{F2C38F}
C14 & Very light brown & Dynamic Projection-Mapping for accessibility \\
\rowcolor[HTML]{7EE4F2}
C15 & Robin egg blue & 3D Projection Mapping \\
\rowcolor[HTML]{EACA52}
C16 & Light mustard & Dynamic Projection-Mapping in stores and virtual try-on \\
\bottomrule
\end{tabularx}
\end{table}

\subsection{Rankings and field orientation}

The author ranking highlighted recurring names such as Daisuke Iwai, Masatoshi Ishikawa, Kosuke Sato, Yoshihiro Watanabe, Bruce H. Thomas, Ramesh Raskar, Mark Billinghurst, Marc Stamminger, Hiromasa Oku, Hiroshi Ishii, Andrew D. Wilson, Naoki Hashimoto, Hrvoje Benko, Shahram Izadi, and Henry Fuchs. Subject terms were dominated by software, computer graphics and computer-aided design, computer vision and pattern recognition, computer science applications, human-computer interaction, electrical and electronic engineering, signal processing, artificial intelligence, general medicine, and control and systems engineering.

These outputs are not substitutes for qualitative reading. Their value is orientation: they help the researcher find influential clusters, identify vocabulary, and decide which communities must be read carefully. In the original doctoral review, this orientation helped clarify why a broad \dpm{} framing was useful at the beginning, while the research could later narrow toward deployable wall-scale projection AR, reproducible \procam{} engineering, multi-application orchestration, and museum-grounded operation.

\section{Discussion}

The \dpm{}/\sar{} case study suggests that network-augmented exploration is useful when a field is broad, hybrid, and fragmented. A purely keyword-based review risks missing neighbouring communities; a purely manual review risks being unstructured; and a purely bibliometric analysis risks producing metrics without a research narrative. \tool{} combines computational assistance with human interpretation.

The method is useful because it makes an informal activity more auditable. Researchers often perform exploratory review work anyway: they search broadly, follow references, identify recurring names, and gradually construct a mental map of a field. \tool{} turns this into a set of artifacts: seed tables, metadata-resolution logs, graph files, community labels, rankings, and browser-inspectable outputs. These artifacts make it easier to justify why a later review was narrowed, why particular subfields were included or excluded, and how a conceptual perimeter was formed.

The method is also useful because it supports interdisciplinary review. In fields like \dpm{}/\sar{}, relevant work appears in computer graphics, human-computer interaction, computer vision, display technology, cultural heritage, manufacturing, and entertainment. A citation graph can reveal communities that are adjacent to the seed terms even when they do not share the exact same vocabulary.

The intended use of \tool{} is therefore methodological rather than authoritative. It helps researchers move from a diffuse literature-search problem to an inspectable and documented map of candidate communities. It does not decide what belongs in the final review, assess study quality, or replace the interpretive responsibility of the researcher.

\section{Threats to validity}

\paragraph{Search-source bias.}
The seed set determines what the graph can discover. If the seed query misses a community, citation expansion may not recover it. This is why Stage 0 explicitly records neighbouring terms and assumptions.

\paragraph{Metadata bias.}
Crossref records vary in completeness. Missing references reduce graph density and can distort centrality. Books, patents, standards, industrial documentation, and non-indexed conference papers may be underrepresented.

\paragraph{Depth bias.}
Depth 1 keeps the corpus interpretable but may miss deeper intellectual ancestry. Higher depths increase recall but risk topic drift. Depth should therefore be reported as a review parameter rather than hidden as an implementation detail.

\paragraph{Community-label subjectivity.}
Modularity classes require human interpretation. Different reviewers may assign different labels to the same cluster. This does not invalidate the method, but it means community labels should be reported as interpretive synthesis rather than objective ground truth.

\paragraph{Metric bias.}
Network centrality is not equivalent to scientific quality. Highly connected papers may be methodological anchors, surveys, or generic background. Conversely, recent or niche contributions may be important despite low degree.

\paragraph{Tooling bias.}
The static web interface emphasizes visibility and inspection, not statistical inference. It should be treated as an exploratory aid and review artifact, not as an automated decision system.

\section{Software and data availability}

The \tool{} repository is available at \url{https://github.com/diogogithub/sota-lens}. It includes the Python package, command-line interface, static web explorer, test suite, manuscript materials, and the bundled \dpm{}/\sar{} case-study corpus. The web interface is available at \url{https://diogogithub.github.io/sota-lens/}. The \tool{} software release v0.1.0 is archived on Zenodo with DOI \href{https://doi.org/10.5281/zenodo.19860899}{10.5281/zenodo.19860899}~\citep{sotalens2026}.

The repository includes the input corpus, graph artifacts, and static explorer used for the proof-of-concept case study. The graph can be regenerated from the bundled data using the command-line pipeline documented in the repository. The generated artifacts include CSV tables, GEXF graph files, browser-ready JSON, and the static web explorer.

\section{AI usage disclosure}

Generative AI assistance was used during repository refactoring, documentation drafting, website copy-editing, test scaffolding, and manuscript editing. The author reviewed, edited, and validated the generated material; selected the scholarly framing; made the core design decisions; checked the repository contents; and remains fully responsible for the accuracy, originality, licensing, and ethical compliance of the submitted work.

\section{Conclusion}

\tool{} provides a lightweight methodology and software toolkit for network-augmented \sota{} reviews. It is intended for the exploratory stage where researchers need to map a field before they can define a narrow systematic review protocol. The proof-of-concept in \dpm{} and \sar{} shows how seed results, metadata resolution, bounded citation expansion, graph construction, community detection, and human labelling can produce an auditable and useful map of a fragmented research landscape. The method does not replace PRISMA, scoping review protocols, systematic mapping, or mature bibliometric tools. Instead, it fills a practical gap between informal narrative exploration and formal evidence synthesis.

\section*{Acknowledgements}

This work emerged from doctoral research on deployable projection-based augmented reality and from a Network Science project at the University of Porto.

\bibliographystyle{plainnat}
\bibliography{references}

\end{document}